# The 'solar model problem' resurrected

Martin Asplund(*), Nicolas Grevesse(+), Manuel Güdel(&) and A. Jacques Sauval($)

*Research School of Astronomy and Astrophysics, Australian National University, Cotter Road, Weston 2611, Australia (martin@mso.anu.edu.au)*

*+ Centre Spatial de Liège and Institut d'Astrophysique et de Géophysique, Université de Liège, B-4000 Liège, Belgium*

*& Paul Scherrer Institut, Wuerenlingen and Villigen, CH-5232 Villigen PSI, Switzerland*

*$ Observatoire Royal de Belgique, 3, Avenue Circulaire, B-1180 Bruxelles, Belgium*

**The new solar composition(ref.1), when applied to compute a model of the Sun, leads to serious disagreement(ref.2) between the predictions of the model and the observations obtained by helioseismology. New measurements of the coronal Ne/O abundance ratio in nearby stars(ref.3,4,5) using X-ray spectra typically find high values of Ne/O=0.4 rather than 0.15 adopted in ref.1 for the Sun. Drake & Testa(ref.3) suggest that this high Ne/O ratio is appropriate also for the Sun, which would bring the solar models back in agreement with the helioseismological observations(ref.2). Here we present arguments why the high Ne/O ratio is unlikely to be applicable to the Sun.**

Stellar coronal abundances often differ from their photospheric counterparts. A full understanding of this fractionation has not yet emerged but it is related to the first ionization potential (FIP) of the elements and magnetic activity in the chromosphere/corona(ref.4,6). For the Sun, low FIP elements like Si, Ca and Fe are enhanced in the corona compared with the photosphere while in highly active stars high FIP elements like C, N and O are over-abundant relative to low FIP elements. As most of the stars presented in refs.(3,4) are much more active than the Sun, we suspect that their high coronal Ne/O is a consequence of the even higher FIP for Ne (21.6 vs 13.6 eV), making Ne more susceptible to fractionation than O; since Ne lacks photospheric measurements this is not possible to test directly however. Better insight to the solar Ne/O should therefore come from low activity dwarf stars. The only such star in ref.3 is Procyon for which other analyses suggest Ne/O=0.22 (ref.4) rather than 0.4 used in ref.(3). Furthermore, missing from ref.(3) is the closest solar analog, alpha Cen A, which has a low coronal Ne/O=0.18 (ref.5), while the inactive dwarf beta Com has Ne/O=0.14 (ref.6). Thus, the sofar studied magnetically inactive stars most similar to the Sun have low coronal Ne/O ratios.

In the solar corona, numerous measurements of the Ne/O ratio, in various types of coronal matter (quiet and active regions with different geometry and levels of activity), using very different techniques (XUV- and gamma-ray spectroscopy, particle collections in the fast and slow solar wind and in various types of solar energetic particles), lead to remarkably constant low values of the Ne/O (0.13-0.19) ratio(ref.7,8). Very few exceptional events show high Ne/O ratios but these are also characterized by other strange isotopic compositions not resembling the photospheric values(ref.8). Such very energetic flares are important for the coronal heating in active stars(ref.9), providing a possible connection with the high coronal Ne/O value observed in these stars.

The solar Ne/O ratio can also be gauged through observations of other nearby Galactic objects, which have similar overall compositions to the Sun. The typically derived Ne/O ratios in hot stars(ref.10), H II regions(ref.11) and planetary nebulae(ref.12) are in very good agreement with the solar value advocated in ref.(1). Finally, both Ne and O are predominantly produced in supernovae type II. Nucleosynthesis predictions for such events suggest a production ratio of Ne/O~0.1-0.2 when integrating over the initial mass function of stars(ref.13). The evidence is thus stacked against the solar Ne/O ratio being as high as 0.4.

The problem of the solar abundance of neon is certainly not yet entirely settled; a survey of more low activity solar analogs should be carried out. Based on the arguments given above, however, we argue that the low Ne/O ratio of 0.15 adopted in ref.1 is the correct value for the present Sun rather than Ne/O~0.4 as proposed in ref.(3). As a consequence, the solar model problem still awaits a solution.